\documentclass[letterpaper, 10 pt, conference]{ieeeconf}  

\IEEEoverridecommandlockouts                              
\usepackage{times}

\usepackage{caption}

\usepackage[siunitx]{circuitikz}

\usepackage{microtype}

\usepackage[hidelinks]{hyperref}

\usepackage{amsfonts}

\usepackage{amsthm}
\usepackage{amssymb}
\usepackage{amsmath}
\usepackage{extarrows}
\usepackage{enumerate}
\usepackage{graphicx}
\usepackage{pgf}
\usepackage{algorithm,algorithmic}
\usetikzlibrary{arrows,automata,fit,matrix}
\usepackage{subfig}
\usepackage{colortbl}

\usepackage{wrapfig}
\usepackage{verbatim}
\usepackage{bm}
\usepackage{bbm}
\usepackage{epstopdf}

\usepackage{mathtools}

\usepackage{dsfont}

\usepackage{booktabs}

\usepackage[off,crop=off]{auto-pst-pdf}

\newtheorem{definition}{Definition}
\newtheorem{example}{Example}

\newtheorem{theorem}{Theorem}

\newcommand{\RE}{\mathbb{R}}

\newcommand{\calH}{\mathcal{H}}
\newcommand{\calK}{\mathcal{K}}

\newcommand{\R}{\mathbb{R}}

\newcommand{\hx}{\hat{x}}

\newcommand{\hA}{\hat{A}}
\newcommand{\hB}{\hat{B}}
\newcommand{\hu}{\hat{u}}

\newcommand{\hF}{\hat{F}}
\newcommand{\hJ}{\hat{J}}

\renewcommand{\hm}{\hat{m}}
\newcommand{\hM}{\hat{M}}

\newcommand{\hR}{\hat{R}}

\newcommand{\hV}{\hat{V}}

\newcommand{\T}{T}

\newcommand{\lb}{m}
\newcommand{\ub}{M}

\newcommand{\hlb}{\hat{\lb}}
\newcommand{\hub}{\hat{\ub}}

\usepackage{array}
\newcolumntype{H}{>{\setbox0=\hbox\bgroup}c<{\egroup}@{}}

\title{\LARGE \bf
Coarse-graining Complex Networks for Control Equivalence
}

\author{Daniele Toller$^{1}$, Mirco Tribastone$^{2}$, Max Tschaikowski$^{1}$, and Andrea Vandin$^{3}$
\thanks{$^{1}$ D. Toller and M. Tschaikowski are with Aalborg University, Denmark}
\thanks{$^{2}$ M. Tribastone is with IMT Lucca, Italy}
\thanks{$^{3}$ A. Vandin is with Sant'Anna Pisa, Italy and DTU Compute, Denmark}
}

\usepackage{numprint}

\begin{document}

\maketitle
\thispagestyle{empty}
\pagestyle{empty}

\begin{abstract}
The ability to control complex networks is of crucial importance across a wide range of applications in natural and engineering sciences. However, issues of both theoretical and numerical nature introduce fundamental limitations to controlling large-scale networks. In this paper, we cope with this problem by introducing a coarse-graining algorithm. It leads to an aggregated network which satisfies control equivalence, i.e., such that the optimal control values for the original network can be exactly recovered from those of the aggregated one. The algorithm is based on a partition refinement method originally devised for systems of ordinary differential equations, here extended and applied to linear dynamics on complex networks. Using a number of benchmarks from the literature we show considerable reductions across a variety of networks from biology, ecology, engineering, and social sciences.        
\end{abstract}


\section{Introduction}\label{sec:intro}

A large variety of complex systems of biological, ecological, social, and technological nature can be conveniently modeled as networks.
Their dynamics can thus be studied as a (typically high-dimensional) system of differential equations where the evolution of each node depends on how it is connected to its neighbors and how it responds to external stimuli. Network control, i.e., the ability to steer the system toward a desired target behavior, is of crucial importance both theoretically and practically~\cite{RevModPhys.88.035006}. This involves two main steps~\cite{8718868}: (i)~determining through which nodes to input a control action; and (ii)~finding the actual values of the inputs.

Network science has made fundamental progress with approaches that identify a minimum subset of nodes to use as drivers in order to achieve control of the system, e.g.,~\cite{Lin1100557,Liu:2011vz,Yuan:2013we,Gao:2014tj}.
The actual computation of a control law is usually a nontrivial task~\cite{2015,doi:10.1063/1.4931570}. For example, using standard results from systems theory~\cite{kailath1980linear}, the controls that minimize energy dissipation can be computed explicitly in networks with linear dynamics. However, this involves the inversion of the controllability matrix, which ultimately gives a cubic dependency of the computational cost with the cardinality of the network~\cite{6762966}.
Furthermore, the realizability of explicit solutions may be hindered in practice by numerical ill-conditioning of the controllability matrix~\cite{PhysRevLett.110.208701}, by physical constraints that preclude the use of arbitrarily large input signals~\cite{6762966,Wang:2017ug}, or control trajectories that evolve far away in the state space~\cite{PhysRevLett.108.218703,doi:10.1126/science.aai7488}. Overall, these issues have been recognized as fundamental limitations to controlling large networks~\cite{PhysRevLett.110.208701,RevModPhys.88.035006}.


Across many branches of science and engineering, coarse graining is a typical strategy to cope with large-scale, intractable problems. Broadly speaking, it can be defined as an approach that maps a model onto a simpler one on which the analysis can be carried out more efficiently. Here we propose a coarse-graining algorithm for complex networks which preserves the control trajectories and the computation of optimal control values.

Although our ultimate goal is the control of networks with nonlinear dynamics, the study of the linear case is a fundamental step in this endeavour. We follow the standard setting~\cite{RevModPhys.88.035006} whereby the linear dynamical system induced by a network with $N$ nodes and adjacency matrix $A \in \RE^{N \times N}$ is given by
\begin{equation}\label{eq:net}
\partial_t x(t) = A x(t) + B  u(t),
\end{equation}
where $x(t) \in \mathbb{R}^N$ is network's state at time $t$, $B \in \mathbb{R}^{N \times K}$ is the input matrix, and $u(t) \in \mathbb{R}^K$ is the control input. The columns of $B$ are given by $K$ different unit vectors and describe the fact that $K$ nodes in the network are \emph{driver nodes}, each with its own distinct control input~\cite{Liu:2011vz}; the remaining $N-K$ nodes, instead, are not controlled. We assume that $u(t)$ is bounded, which can thus take into account constraints for physical realizability~\cite{BEMPORAD20023,6762966,Wang:2017ug}.


In this context, the synthesis of a control law can be phrased as a problem of minimizing a cost function using $u(t)$ as the set of decision variables and~\eqref{eq:net} as the constraints. We consider a rather general cost function in the form
\begin{equation}\label{eq:cost}
J(x[0],u,T) = F(x(T)) + \int_0^{T} \big[ S(t, x(t)) + Q(t, u(t)) \big] dt ,
\end{equation}
where the function $F$, called the final cost, may be used to steer the system toward a given target state at final time $T$; the integrand $R = S + Q$ is called the running cost. Note that the above integral is computed along the trajectory $x(t)$ determined by the initial condition $x[0]$, and the control input $u$. This covers the classical setting where only the energy of the control is to be minimized~\cite{RevModPhys.88.035006}; the term $S$ allows for objectives where one is interested in driving the system toward a desired trajectory~\cite{RevModPhys.88.035006}, possibly in a targeted fashion where only a subset of the state space, and not full control, is of interest, for instance \cite{Gao:2014tj,Klickstein:2017wv}.

Our method partitions the nodes of a network into $n \leq N$ macro-nodes and builds a coarse-grained adjacency matrix $\hat{A} \in \RE^{n \times n}$ and input matrix $\hat{B} \in \mathbb{R}^{n \times k}$, where $k \leq K$ is the number of macro-inputs. The state of the coarse-grained network $\hat{x}(t) \in \mathbb{R}^n$ is the solution of the system
\begin{equation}\label{eq:cg.net}
\partial_t \hx(t) = \hA \hx(t) + \hB \hu(t),
\end{equation}
and $\hu(t) \in \mathbb{R}^{k}$ is the control macro-input.
The partition is such that each macro-state of the coarse-grained network exactly preserves the sum of the values of the corresponding states in the original network whenever each macro-input is equal to the sum of the corresponding original input values.


The preservation of the dynamics up to sums of variables has been already established for linear and nonlinear systems of ordinary differential equations (ODEs) using the notion of forward equivalence~\cite{pnas17}. Importantly, here we prove \emph{control equivalence}, a property that guarantees the preservation of optimal costs. For this, we assume that the cost functional in~\eqref{eq:cost} can be equivalently rewritten in terms of a cost functional $\hat{J}(\hat{x}, \hat{u},T)$ that depends on the coarse-grained model. Under these conditions, the solution of the control problem for minimizing $\hat{J}(\hat{x}, \hat{u},T)$ with the constraints given from~\eqref{eq:cg.net} is such that one can recover  a mapping from the optimal macro-inputs $\hat{u}^\ast(t)$ to optimal original inputs $u^\ast(t)$ that give rise to the optimal trajectory in~\eqref{eq:net}. With this, the analysis of the original control problem can be entirely circumvented.

The crux of our method is that the coarse-graining can be found by computing the forward equivalence for the ODEs system $\partial_t x(t) = A x(t)$. Importantly, it can be performed using the partition-refinement algorithm defined in~\cite{pnas17}. The algorithm provides the reduction as the coarsest partition that refines an initial candidate partition, iteratively splitting its blocks until certain criteria are met~\cite{DBLP:conf/popl/CardelliTTV16,pnas17}.
As a straightforward consequence of this fact, from the  computational complexity analysis in~\cite{pnas17} we obtain that control equivalence can be computed in $O(E \log N)$ time, where $E$ is the number of non-zero entries in $A$.

The initial partition may be arbitrarily chosen. Similarly to the other applications of partition refinement~\cite{doi:10.1137/0216062,Kanellakis:1983:CEF:800221.806724,Derisavi2003309,DBLP:conf/tacas/ValmariF10}, this freedom permits certain state variables to be kept observable in the coarse-grained model. For example, let us assume that the original cost function $J(x,u,T)$ only concerns a given state component $x_i$; then, one can ensure by construction that the coarse-grained cost function $\hat{J}(\hat{x},\hat{u},T)$ can be written in terms of $x_i$ by placing it in an initial singleton block for the partition-refinement algorithm.

\paragraph*{Results} Our main technical contribution is a characterization result. Here, we prove that the reduction provided by our reduction algorithm is maximal in the sense that there exists no coarser refinement of a given initial partition which yields an optimality-preserving reduction. Additionally to the technical result, we conduct a large-scale evaluation of our framework on networks from public repositories. Here, the main finding is that networks that are controlled by a minimal set of driver nodes, as computed through~\cite{Liu:2011vz}, often allow for further substantial optimality-preserving reductions.

\paragraph*{Further related work} Aggregation of dynamical systems has been extensively studied in the control theory literature since a long time starting from Aoki~\cite{1098900}; while it can be related to control equivalence, a method for computing is not provided. More closely related is the bisimulation/abstraction of control systems~\cite{bisimulation_lin_sys_Schaft,DBLP:journals/tac/PappasS02,DBLP:journals/automatica/TabuadaP03} where the largest bisimulation gives rise to the smallest reduced control system which coincides with the original one up to a given observation map. The main similarity with the bisimulation approaches is that control equivalence computes such an observation map, for which, in particular, the reduced system can be shown to be a so-called \emph{consistent implementable abstraction} of the original one~\cite{DBLP:journals/tac/PappasLS00}. 

To the best of our knowledge, a (non-unique dimension-based) computation of observation maps has been addressed in the case with unbounded control domains~\cite{DBLP:journals/tac/PappasLS00}. Being described by systems of linear equations, it relies on matrix transformations and has thus a cubic time complexity in the number of nodes. Control equivalence, instead, is described by structural properties~\cite{DBLP:conf/tacas/CardelliTTV16,tognazzi2018differential} that are used in the computation of graph isomorphisms~\cite[Definition 5]{hartke2009mckay}. Hence, while considering a smaller class of observation maps than bisimulation, control equivalence avoids matrix transformations and enjoys a quasilinear time complexity in the number of edges; moreover, it considers bounded control domains. A less closely related work is that of balanced truncation and its extensions, see for instance~\cite[Chapter 9]{doi:10.1137/1.9780898718713} and~\cite{9112314}. This is because the respective reductions are not exact and the control domains are unbounded; in addition, they have cubic time complexity~\cite{doi:10.1137/1.9780898718713}. Bisimulation approaches have been also studied for nonlinear systems~\cite{DBLP:journals/tcs/CardelliTTV19a,DBLP:conf/qest/CardelliTTV18,bisimulation_lin_sys_Schaft,DBLP:journals/tac/PappasS02}, are however computationally more demanding than the respective linear counterparts.


\paragraph*{Notation} We denote by $N$ and $K$, respectively, the number of nodes and driver nodes in the original network; similarly, $n \leq N$ and $k \leq K$ have the respective meaning in the reduced network. The respective dynamics of the original and the reduced networks are described by Eq.~\ref{eq:net} and Eq.~\ref{eq:cg.net}, where $x,\hx$ are state vectors, while $u,\hu$ are control vectors. Letters $i,j$ denote node indices, $l,l'$ refer to control inputs, $\calH,\calH'$ refer to partitions of $\{1,\ldots,N\}$, while $H,H'$ to partition blocks. A partition $\calH$ is said to refine a partition $\calH'$ if every block $H' \in \calH'$ can be expressed in terms of a disjoint union of blocks from $\calH$. 
Note that $\calH$ is a trivial refinement 
of itself. Likewise, $\{ \{1,\ldots,N\} \}$ and $\{ \{i\} \mid 1 \leq i \leq N \}$ are considered to be trivial partitions because the former aggregates all nodes to a single one, while the latter does not aggregate anything.


\section{Control Equivalence}

For two vectors $\lb\in \R^{K}$ and $\ub\in \R^{K}$ with respective $l$-th entries $\lb_l \leq \ub_l$ for every $l = 1, \ldots, K$, consider the real interval $[\lb_l; \ub_l]\subseteq \R$ and the $K$-dimensional cube 
\[
[\lb;\ub] = \prod_{l = 1}^K [\lb_l; \ub_l] \subseteq \R^K.
\]
We always consider control inputs $u(\cdot) \in [\lb;\ub]$, by which we mean that $u \colon \mathbb{R} \to [\lb;\ub]$ is a measurable function of time taking values in $[\lb;\ub]$.

\begin{definition}[Optimal costs]
Given a matrix $A \in \RE^{N \times N}$ and a measurable control input $u(\cdot) \in [m;M]$, we denote the value functional at time $\T \in \R$ when starting at $x[0] \in \R^{N}$ by
\begin{align*}
J_{u}(x[0], \T) & = F \big( x^{u} (\T) \big) + \int_{0}^\T R\big( t, x^{u} (t), u(t) \big) dt,
\end{align*}
where $x^{u}$ is the solution of~\eqref{eq:net}, with initial condition $x[0]$, while $F$ is the final cost function, and $R = S + Q$ is the running cost function. 
With this, the optimal costs at time $\T$ when starting at $x[0] \in \R^{N}$ are given by the value functions:
\begin{align*}
V^{\inf} (x[0], \T) &= \inf \{ J_{u}(x[0], \T) \mid u: \R \to [\lb;\ub] \}, \\
V^{\sup} (x[0], \T) &= \sup \{ J_{u}(x[0], \T) \mid u: \R \to [\lb;\ub] \},
\end{align*}
where the control inputs $u(\cdot)$ are measurable.
\end{definition}

\begin{figure}[pt!]
\centering
 \begin{tikzpicture}[->,>=stealth,shorten >=1pt,auto,semithick]
    \matrix[row sep=0.77cm, column sep=0.77cm]
    {
    \pgfmatrixnextcell \pgfmatrixnextcell \node[circle](1){$1$}; \pgfmatrixnextcell \pgfmatrixnextcell \\
    \node[circle](4){$[1;2] \ni u_1$}; \pgfmatrixnextcell \node[circle](2){$2$}; \pgfmatrixnextcell \pgfmatrixnextcell \node[circle](3){$3$}; \pgfmatrixnextcell \node[circle](5){$u_2 \in [3;4]$}; \\
    };
    \path (2) edge[bend right] node {$\tfrac{1}{2}$} (1)
        (3) edge [bend left] node [swap] {$\tfrac{1}{2}$} (1)
        (1) edge [bend right] node [swap] {$\tfrac{1}{4}$} (2)
        (1) edge [bend left] node {$\tfrac{2}{4}$} (3)
        (4) edge [dashed] node {$ $} (2)
        (5) edge [dashed] node {$ $} (3);
    \end{tikzpicture}
    \caption{A simple controlled network which will be used as a running example. Solid lines describe entries of the adjacency matrix, while dashed lines signify control inputs.}\label{fig:re}
\end{figure}
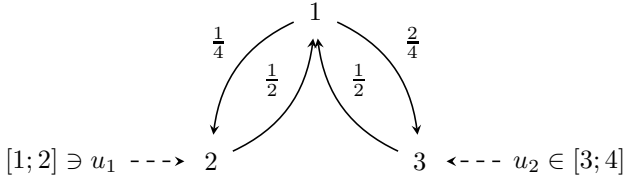

In the following, we shall use the simple network provided in Fig.~\ref{fig:re} to explain the main concepts.

\begin{example}\label{ex:1}
Eq.~\ref{eq:net} of the network from Fig.~\ref{fig:re} is given by
\begin{equation*}
\begin{pmatrix}
\partial_t x_1 \\
\partial_t x_2 \\
\partial_t x_3
\end{pmatrix} =
\underbrace{
\begin{pmatrix}
0 & \tfrac{1}{2} & \tfrac{1}{2} \\
\tfrac{1}{4} & 0 & 0 \\
\tfrac{1}{2} & 0 & 0
\end{pmatrix}
}_{A =}
\begin{pmatrix}
x_1 \\
x_2 \\
x_3
\end{pmatrix}
+
\underbrace{
\begin{pmatrix}
0 & 0 \\
1 & 0 \\
0 & 1
\end{pmatrix}
}_{B =}
\begin{pmatrix}
u_1 \\
u_2
\end{pmatrix} .
\end{equation*}
\end{example}

As next, we introduce control equivalence. For a set $H$, we denote its cardinality by $|H|$.

\begin{definition}[Control Equivalence]
Consider the system of ODEs in \eqref{eq:net}. A control equivalence of this system is a partition $\calH$ whose aggregation matrix $L = L(\calH)$ and the corresponding right inverse $\bar{L} = \bar{L}(\calH)$ satisfy $LA  = LA\bar{L}L$. Specifically, $L \in \RE^{n \times N}$ and $\bar{L} \in \RE^{N \times n}$ are such that
\begin{itemize}
    \item The $i$-th row of $L$ encodes the $i$-th block of $\calH$ by setting $L_{i,j} = 1$ if $j \in H_i$ and zero otherwise;
    \item The $i$-th column of $\bar{L}$ encodes the $i$-th block of $\calH$ by setting $\bar{L}_{j,i} = 1/|H_i|$ if $j \in H_i$ and zero otherwise.
\end{itemize}
\end{definition}

When the partition $\calH$ is clear from the context, we omit the dependence on $\calH$ in $L(\calH)$ and $\bar{L}(\calH)$.

\begin{example}\label{ex:2}
In the case of the network from Fig.~\ref{fig:re}, it is easy to see that the trivial partition $\{ \{1,2,3\} \}$ is not a control equivalence, which corresponds to the impossibility of aggregating all nodes together. 

On the other hand, the partition $\calH = \{ H_1, H_2 \} $ $= \{ \{2,3\}, \{1\} \}$ is a control equivalence (intuitively: it is possible to aggregate the nodes $2$ and $3$). 
This can be seen by checking that the matrices
\begin{align*}
L & =
\begin{pmatrix}
0 & 1 & 1 \\
1 & 0 & 0
\end{pmatrix}
& \text{and} & &
\bar{L} & =
\begin{pmatrix}
0 & 1 \\
\tfrac{1}{2} & 0 \\
\tfrac{1}{2} & 0
\end{pmatrix}
\end{align*}
satisfy $LA  = LA\bar{L}L$.
\end{example}

Given a partition $\calH = \{H_1,\ldots,H_n\}$ of nodes $\{1,\ldots,N\}$, we always assume without loss of generality that its first $k$ blocks $H_1,\ldots,H_k$ contain all at least one driver node, while the remaining blocks $H_{k+1},\ldots,H_n$ have no driver nodes.

\begin{definition}[Reduced System]\label{def:red:sys}
Consider the system of ODEs in \eqref{eq:net}. 
Then, the reduced system underlying a partition $\calH$ is given by~\eqref{eq:cg.net} where
\begin{itemize}
    \item $\hA = L A \bar{L}$;
    \item For any $l \leq k$, the $l$-th column of $\hB \in \RE^{n \times k}$ is the $l$-th unit vector in $\RE^n$;
    \item For any $l \leq k$, the $l$-th control satisfies $\hu_l(\cdot) \in [\hat{m}_l ; \hat{M}_l]$ with
    \begin{align*}
    \hat{m}_l & = \sum_{l' \in \calK(H_l)} m_{l'} & \text{and} & & \hat{M}_l & = \sum_{l' \in \calK(H_l)} M_{l'} ,
    \end{align*}
    where $\calK(H_l) \subseteq \{1,\ldots,K\}$ are the control indices that steer the driver nodes contained in block $H_l$.
\end{itemize}
\end{definition}

Note that the sets $\calK(H_1), \calK(H_2), \ldots$ form a partition of $\{1,\ldots,K\}$ (i.e., of the driver nodes). We used this fact to well-define the bounds $\hat{m}$ and $\hat{M}$ for the macro-inputs.

\begin{example}\label{ex:3}
Continuing Example~\ref{ex:2}, the reduced system is
\begin{equation*}
\begin{pmatrix}
\partial_t \hx_1 \\
\partial_t \hx_2 \\
\end{pmatrix} =
\underbrace{
\begin{pmatrix}
0 & \tfrac{3}{4} \\
\tfrac{1}{2} & 0 \\
\end{pmatrix}
}_{\hA =}
\begin{pmatrix}
\hx_1 \\
\hx_2
\end{pmatrix}
+
\underbrace{
\begin{pmatrix}
1 \\
0
\end{pmatrix}
}_{\hB =}
\begin{pmatrix}
\hu_1
\end{pmatrix} ,
\end{equation*}
where $\hu_1(\cdot) = u_1(\cdot) + u_2(\cdot) \in [4;6]$, while $\hx_1 = x_2 + x_3$ and $\hx_2 = x_1$. We note, in particular, that $\hx_2 = x_1$ is not controlled, i.e., is not a driver node.
\end{example}

In order to associate a cost to a reduced system, the final and running costs have to satisfy an additional property. More specifically, for a partition of nodes $\calH$, we call the final cost $F$ and running cost $R$ \emph{constant on $\calH$} if $F(x) = F(y)$ and $R(t,x,u) = R(t,y,v)$ for all $x,y \in \RE^N$, and $u,v \in \RE^K$ 
that satisfy $Lx = Ly$ and $LBu = LBv$, for $L = L(\calH)$. 
In such a case, we study the well-defined reduced costs $\hR$ and $\hF$ for the reduced system defined for $\hx \in \RE^n$ and $\hu \in [\hm ; \hM]$ via
\begin{align*}
\hR(t,\hx,\hu) & := R(t,x,u) & \text{and} & & \hF(\hx) & := F(x) ,
\end{align*}
for arbitrary $x$ and $u$ such that $\hx = Lx$ and $\hB \hu = LBu$ 
(equivalently, $\hx_i = \sum_{j \in H_i} x_j$ and $\hu_l = \sum_{l' \in \calK(H_l)} u_{l'}$ for all $1 \leq i \leq n$ and $1 \leq l \leq k$).

\begin{example}\label{ex:4}
We continue Example~\ref{ex:3} by picking as running cost and final cost, respectively, $R = 0$ and $F(x) = x_2 + x_3$. Then, $R$ and $F$ are constant on $\calH = \{ \{2,3\}, \{1\} \}$ and we obtain $\hat{R}(\hx) = 0$ and $\hat{F}(\hx) = \hx_1$. 
With these cost functions, the value functionals have the form
\[
J_u(x[0],T) = F (x^u(T)) = x_2^u(T) + x_3^u(T). 
\] 
\end{example}

\begin{definition}[Optimal cost of Reduced System]
Let $\calH$ be an arbitrary partition, and assume that $F$ and $R$ are, respectively, final and running costs that are constant on $\calH$. Then, for a control input $\hu(\cdot) \in [\hm ; \hM]$, the value functional at time $T \in \R$ when starting at $\hx[0] \in \R^n$ is given by
\begin{align*}
\hJ_{\hu}(\hx[0], \T) & = \hF \big( \hx^{\hu} (\T) \big) + \int_{0}^\T \hR\big( t, \hx^{\hu} (t), \hu(t) \big) dt,
\end{align*}
where $\hx^{\hu}$ is the solution of \eqref{eq:cg.net}, with initial condition $\hx[0]$. With this, the optimal costs of the reduced system \eqref{eq:cg.net} at time $\T$ when starting at $\hx[0] \in \R^n$ are given by the reduced value functions:
\begin{align*}
\hV^{\inf} (\hx[0], \T) &= \inf \{ \hJ_{\hu}(\hx[0], \T) \mid \hu: \R \to [\hlb;\hub] \}, \\
\hV^{\sup} (\hx[0], \T) &= \sup \{ \hJ_{\hu}(\hx[0], \T) \mid \hu: \R \to [\hlb;\hub] \},
\end{align*}
where the control inputs $\hu(\cdot)$ are measurable.
\end{definition}

We are now in a position to state our first result which is key in proving that the original and the reduced control system admit identical optimal costs if and only if $\calH$ is a control equivalence.

\begin{theorem}[Optimality-preservation]\label{thm:main1}
Let $\calH$ be a control equivalence of the system \eqref{eq:net} with final and running costs that are constant on $\calH$.
\begin{enumerate}
    \item For any control of the original system $u : \RE \to [\lb ; \ub]$, there exists a control of the reduced system $\hu: \R \to [\hlb;\hub]$ that satisfies
    \[
    J_{u}(x[0],\T) = \hJ_{\hu}(\hx[0],\T)
    \]
    for every $\T \in \R$ and every $x[0] \in \R^{N}$, where $\hx[0] \in \RE^n$ is given by $\hx[0] = Lx[0]$. 

    \item Conversely, for any control of the reduced system $\hu: \R \to [\hlb;\hub]$, there exists a control of the original system $u : \RE \to [\lb ; \ub]$ that satisfies
    \[
    J_{u}(x[0],\T) = \hJ_{\hu}(\hx[0],\T)
    \]
    for every $\T \in \R$ and every $x[0] \in \R^{N}$, where $\hx[0] \in \RE^n$ is given by $\hx[0] = Lx[0]$. Additionally, the control $u$ can be chosen as
    \[
    u_{l'}(\cdot) =
    \begin{cases}
        \displaystyle{ m_{l'} + \frac{ M_{l'} - m_{l'} }{ \hM_{l}-\hm_{l} } ( \hu_{l}(\cdot) - \hm_{l} ) }
                        & , \ \text{if } \hm_{l} < \hM_{l} \\
        m_{l'} = M_{l'}   & , \ \text{otherwise}
    \end{cases}
    \]
    for every $l \in \{1,\ldots,k\}$ and $l' \in \calK(H_l)$.
\end{enumerate}
\end{theorem}

Armed with Theorem~\ref{thm:main1}, we infer our main technical result.

\begin{theorem}[Characterization of Optimality-preservation]\label{thm:main2}
Consider the system in \eqref{eq:net}, and let $\calH$ be a partition of the nodes. Then, $\calH$ is a control equivalence if and only if
\begin{align*}
V^{\inf}(x[0],\T) &= \hat{V}^{\inf}(\hx[0], \T), \\
V^{\sup}(x[0],\T) &= \hat{V}^{\sup}(\hx[0], \T)
\end{align*}
for all final and running costs that are constant on $\calH$, for every $\T \in \R$, and every $x[0] \in \R^{N}$, where $\hx[0] \in \RE^n$ is determined by $\hx[0] = Lx[0]$. 
\end{theorem}

\begin{example}\label{ex:5}
In the case of our running example, and with the cost functions as in Example \ref{ex:4}, we obtain:
\begin{itemize}
\item 
For any $u_1(\cdot) \in [1;2]$ and $u_2(\cdot) \in [3;4]$, there exists $\hu(\cdot) \in [4;6]$ such that for any $x[0] \in \RE^3$ and $\hx[0] = Lx[0]$,  for every $T$ we have that 
\[
x_2^u(T) + x_3^u(T) = \hat{x}_1^{ \hat{u} }(T).
\]
\item 
For any $\hu_1(\cdot) \in [4;6]$, there exist $u_1(\cdot) \in [1;2]$ and $u_2(\cdot) \in [3;4]$ such that for any $\hx[0] \in \RE^2$ and $x[0] \in \RE^3$ with $\hx[0] = Lx[0]$, we have that 
\[
x_2^u(T) + x_3^u(T) = \hat{x}_1^{ \hat{u} }(T).
\]
Specifically, we can pick
\begin{align*}
u_1(t) & = 1 + \tfrac{1}{2}(\hu_1(t) - 4), &
u_2(t) & = 3 + \tfrac{1}{2}(\hu_1(t) - 4).
\end{align*}
\end{itemize}
\end{example}

We end the section by stating that control equivalence can be computed in quasilinear time in the number of edges. As the trivial partition made of singletons is a control equivalence, the refining procedure in the next theorem is guaranteed to terminate.

\begin{theorem}[Computation of Control Equivalence]\label{thm:main3}
Consider system~\eqref{eq:net} and let $\calH'$ be a partition of the nodes. Then the coarsest control equivalence refining $\calH'$, denoted by $\calH$, and the underlying reduced system $\partial_t \hx(t) = \hA \hx(t) + \hB \hu(t)$, can be computed in at most $O(E \log N)$ steps, where $E$ is the number of non-zero entries of $A \in \RE^{N \times N}$.
\end{theorem}

\begin{example}\label{re:6}
If we apply the partition refinement from~\cite{pnas17} on the partition $\calH' = \{ \{1,2,3\} \}$, the algorithm will refine the block $\{1,2,3\}$ to the blocks $\{1\}, \{2,3\}$, thus providing us with the previously seen partition $\calH = \{ \{1\}, \{2,3\} \}$.
\end{example}

We remark that the partition $\calH'$ in Theorem~\ref{thm:main3} can be used to steer the search for control equivalences. Indeed, if one is interested in finding a reduction which does not aggregate node $1$ with any other node, one could pick $\calH' = \{ \{1\}, \{2,3\} \}$. With this choice, the partition refinement algorithm either splits the block $\{2,3\}$ into $\{2\}, \{3\}$ or not. From the foregoing discussion, we know that the latter will be the case because $\{ \{1\}, \{2,3\} \}$ is a control equivalence. On the contrary, if one would use $\calH' = \{ \{1,2\}, \{3\} \}$ (say, because one wants to obtain a reduced model which does not aggregate node $x_3$ with any other node), then the algorithm would split the block $\{1,2\}$ and return the (trivial) control equivalence $\{ \{1\}, \{2\}, \{3\} \}$ which does not aggregate any node.

\section{Evaluation}

\begin{table*}
\centering
\begin{tabular}{
c 
c 
H 
H 
H 
r 
r 
r 
r 
r 
r 
r 
}
\toprule
\emph{Name} & \emph{Type} & Source & Network Type & Model Type & \multicolumn{1}{c}{$N$} & \multicolumn{1}{c}{$n$} & \multicolumn{1}{c}{$n/N$} &
\multicolumn{1}{c}{$K$} & \multicolumn{1}{c}{$k$}  & \multicolumn{1}{c}{$k/K$} & \multicolumn{1}{c}{$K/N$} \\
\midrule
Kohonen & CITATION & SparseMatrixCollection~\cite{10.1145/2049662.2049663} & DIRECTED & LINEAR & 3773 & 2652 & 70.29 & 2114 & 1123 & 53.12 & 56.03 \\
subelj\_cora & CITATION & KONECT~\cite{konect} & DIRECTED & LINEAR & 23167 & 9254 & 39.94 & 10210 & 281 & 2.75 & 44.07 \\
CitHepTh & CITATION & OTHER & DIRECTED & LINEAR & 27771 & 19764 & 71.17 & 5994 & 650 & 10.84 & 21.58 \\
CitHepPh & CITATION & OTHER & DIRECTED & LINEAR & 34547 & 24936 & 72.18 & 8030 & 802 & 9.99 & 23.24 \\
\midrule
s208st & ELECTRONIC CIRCUITS & BARABASI~\cite{Liu:2011vz} & DIRECTED & LINEAR & 123 & 105 & 85.37 & 29 & 15 & 51.72 & 23.58 \\
s420st & ELECTRONIC CIRCUITS & BARABASI & DIRECTED & LINEAR & 253 & 214 & 84.58 & 59 & 29 & 49.15 & 23.32 \\
s838st & ELECTRONIC CIRCUITS & BARABASI & DIRECTED & LINEAR & 513 & 438 & 85.38 & 119 & 60 & 50.42 & 23.20 \\
\midrule
seagrass & FOOD WEB & BARABASI & DIRECTED & LINEAR & 50 & 43 & 86.00 & 13 & 8 & 61.54 & 26.00 \\
ythan & FOOD WEB & BARABASI & DIRECTED & LINEAR & 136 & 89 & 65.44 & 69 & 24 & 34.78 & 50.74 \\
grassland & FOOD WEB & BARABASI & DIRECTED & LINEAR & 89 & 31 & 34.83 & 46 & 10 & 21.74 & 51.69 \\
littlerock & FOOD WEB & BARABASI & DIRECTED & LINEAR & 184 & 54 & 29.35 & 99 & 4 & 4.04 & 53.80 \\
%
\midrule
maayan-faa & INFRASTRUCTURE & KONECT & DIRECTED & LINEAR & 1227 & 1027 & 83.70 & 363 & 215 & 59.23 & 29.58 \\
opsahl-open-flights & INFRASTRUCTURE & KONECT & DIRECTED & LINEAR & 2940 & 2422 & 82.38 & 872 & 469 & 53.78 & 29.66 \\
US-powergrid-4941 & INFRASTRUCTURE & OTHER & DIRECTED & LINEAR & 4942 & 4688 & 94.86 & 575 & 445 & 77.39 & 11.63\\
tntp-Chicago-Regional & INFRASTRUCTURE & KONECT & DIRECTED & LINEAR & 12980 & 12969 & 99.92 & 215 & 215 & 100.00 & 1.66 \\
%
\midrule
p2p-Gnutella08 & INTERNET & KONECT & DIRECTED & LINEAR & 6302 & 5273 & 83.67 & 4106 & 3116 & 75.89 & 65.15 \\
p2p-Gnutella06 & INTERNET & BARABASI & DIRECTED & LINEAR & 8718 & 7737 & 88.75 & 5033 & 4085 & 81.16 & 57.73 \\
p2p-Gnutella05 & INTERNET & BARABASI & DIRECTED & LINEAR & 8847 & 7800 & 88.17 & 5111 & 4115 & 80.51 & 57.77 \\
p2p-Gnutella04 & INTERNET & BARABASI & DIRECTED & LINEAR & 10877 & 9872 & 90.76 & 6004 & 5024 & 83.68 & 55.20 \\
p2p-Gnutella-24A & INTERNET & SparseMatrixCollection & DIRECTED & LINEAR & 26519 & 20273 & 76.45 & 18965 & 12875 & 67.89 & 71.51 \\
%
\midrule
Consulting & INTRA-ORGANIZATIONAL & BARABASI & DIRECTED & LINEAR & 47 & 47 & 100.00 & 2 & 2 & 100.00 & 4.26 \\
Manufacturing & INTRA-ORGANIZATIONAL & BARABASI & DIRECTED & LINEAR & 78 & 74 & 94.87 & 1 & 1 & 100.00 & 1.28 \\
\midrule
rhesusbrain1 & BRAIN CONNECTIONS & BRAIN CONNECTIONS & DIRECTED & LINEAR & 243 & 229 & 94.24 & 242 & 228 & 94.21 & 99.59 \\
celegans-neuronal & NEURONAL & BARABASI & DIRECTED & LINEAR & 298 & 264 & 88.59 & 49 & 17 & 34.69 & 16.44 \\
coliInter-NoAutoReg & METABOLIC & BARABASI & DIRECTED & LINEAR & 420 & 40 & 9.52 & 314 & 2 & 0.64 & 74.76 \\
coliInter-full & METABOLIC & BARABASI & DIRECTED & LINEAR & 425 & 49 & 11.53 & 309 & 2 & 0.65 & 72.71 \\
TRNYeast2-Costanzo & REGULATORY & BARABASI & DIRECTED & LINEAR & 689 & 92 & 13.35 & 565 & 43 & 7.61 & 82.00 \\
maayan-Stelzl & PROTEIN-PROTEIN & KONECT & DIRECTED & LINEAR & 1707 & 1276 & 74.75 & 765 & 420 & 54.90 & 44.82 \\
TRNYeast1-Balaji & REGULATORY & BARABASI & DIRECTED & LINEAR & 4441 & 1983 & 44.65 & 4282 & 1829 & 42.71 & 96.42 \\
yeast2017-full & REGULATORY & OTHER & DIRECTED & LINEAR & 6854 & 6645 & 96.95 & 6648 & 6439 & 96.86 & 96.99 \\
yeast2019-full & REGULATORY & OTHER & DIRECTED & LINEAR & 6887 & 6705 & 97.36 & 6670 & 6488 & 97.27 & 96.85 \\
ownership & REGULATORY & BARABASI & DIRECTED & LINEAR & 7254 & 360 & 4.96 & 5950 & 152 & 2.55 & 82.02 \\
\midrule
moreno-innovation & SOCIAL & KONECT & DIRECTED & LINEAR & 242 & 214 & 88.43 & 29 & 7 & 24.14 & 11.98 \\
CSphdA & SOCIAL & SparseMatrixCollection & DIRECTED & LINEAR & 1883 & 241 & 12.80 & 1176 & 91 & 7.74 & 62.45 \\
ego-twitter & SOCIAL & KONECT & DIRECTED & LINEAR & 23371 & 2692 & 11.52 & 22432 & 2202 & 9.82 & 95.98 \\
ego-gplus & SOCIAL & KONECT & DIRECTED & LINEAR & 23629 & 2438 & 10.32 & 23497 & 2339 & 9.95 & 99.44 \\
Epinions & SOCIAL & BARABASI & DIRECTED & LINEAR & 75880 & 37651 & 49.62 & 41627 & 7450 & 17.90 & 54.86 \\
Slashdot & SOCIAL & BARABASI & DIRECTED & LINEAR & 82169 & 65436 & 79.64 & 3737 & 3680 & 98.47 & 4.55 \\
\midrule
linux & SOURCE CODE & KONECT & DIRECTED & LINEAR & 30838 & 3767 & 12.22 & 20049 & 260 & 1.30 & 65.01 \\
%
\midrule
libre-film-trust & TRUST & KONECT & DIRECTED & LINEAR & 875 & 425 & 48.57 & 359 & 93 & 25.91 & 41.03 \\
WikiVote & TRUST & BARABASI & DIRECTED & LINEAR & 7116 & 2321 & 32.62 & 4736 & 3 & 0.06 & 66.55 \\
librec-ciao-dvd & TRUST & KONECT & DIRECTED & LINEAR & 4659 & 3059 & 65.66 & 3247 & 1748 & 53.83 & 69.69 \\
\midrule
moreno\_blogs & WWW & KONECT & DIRECTED & LINEAR & 1225 & 889 & 72.57 & 436 & 123 & 28.21 & 35.59 \\
dimacs10\_polblogs & WWW & KONECT & DIRECTED & LINEAR & 1225 & 1177 & 96.08 & 126 & 81 & 64.29 & 10.29 \\
wikipedia\_link\_gag & WWW & KONECT & DIRECTED & LINEAR & 2930 & 1097 & 37.44 & 1185 & 52 & 4.39 & 40.44 \\
EPAA & WWW & SparseMatrixCollection & DIRECTED & LINEAR & 4272 & 680 & 15.92 & 3285 & 161 & 4.90 & 76.90 \\
wikipedia\_link\_csb & WWW & KONECT & DIRECTED & LINEAR & 5562 & 2274 & 40.88 & 2971 & 233 & 7.84 & 53.42 \\
CaliforniaA & WWW & SparseMatrixCollection & DIRECTED & LINEAR & 6176 & 1045 & 16.92 & 4489 & 96 & 2.14 & 72.68 \\
wikipedia\_link\_mi & WWW & KONECT & DIRECTED & LINEAR & 7997 & 2185 & 27.32 & 4825 & 153 & 3.17 & 60.34 \\
wbcsstanfordA & WWW & SparseMatrixCollection & DIRECTED & LINEAR & 9436 & 5326 & 56.44 & 3653 & 794 & 21.74 & 38.71 \\
cfinder\_google & WWW & KONECT & DIRECTED & LINEAR & 15764 & 8678 & 55.05 & 5313 & 1487 & 27.99 & 33.70 \\
wikipedia\_link\_bat\_smg & WWW & KONECT & DIRECTED & LINEAR & 21901 & 6845 & 31.25 & 14047 & 891 & 6.34 & 64.14 \\
\bottomrule
\end{tabular}
\caption{Reductions of published networks taken from~\cite{konect,Liu:2011vz,10.1145/2049662.2049663}. $N$ and $n$ is, respectively, the size of the original and reduced network. $K$ is the minimal number of driver nodes of the original network, computed using~\cite{Liu:2011vz}, while $k$ is the number of driver nodes of the reduced network obtained via Theorem~\ref{thm:main3} and Definition~\ref{def:red:sys}. All reductions are optimality-preserving by Theorem~\ref{thm:main1} and~\ref{thm:main2}.
}\label{tab:results}
\end{table*}
We applied control equivalence to benchmark real-world directed networks studied in the literature~\cite{konect,Liu:2011vz,10.1145/2049662.2049663}. They are taken from several domains including biology,  ecology, engineering, online social networks, and the web with sizes ranging from 47 to 82169 nodes. The chosen networks show a wide spectrum of controllability ranging from 0.91\% to 99.59\% of driver nodes in the network as provided by the algorithm from~\cite{Liu:2011vz} which yields a minimal set of driver nodes for network controllability.
We considered initial partitions that separated the set of driver nodes from the rest of the network. In this way one can identify the macro-nodes in the coarse-grained network that represent original driver nodes.

Overall, these benchmark networks show a full range of reducibility by control equivalence measured as $\rho = n/N$, i.e., the  ratio between the number $n$ of macro-nodes and the number $N$ of original nodes. This ranges from $0.10$ in the case of metabolic networks to $0.999$ in a road network (Table~\ref{tab:results}). There is a high correlation ($0.86$) between the overall reduction and the reduction of driver nodes, indicated by $\rho_D = k/K$, i.e., the percentage ratio between the number $k$ of macro driver nodes and the number $K$ of original driver nodes. The former is obtained via Definition~\ref{def:red:sys} and by initializing the reduction algorithm from Theorem~\ref{thm:main3} with an initial partition containing two blocks, one containing all driver nodes and its complement (as mentioned earlier, the set of driver nodes is computed using~\cite{Liu:2011vz} and is minimal in size). With this in place, we note that in some networks, very small values of $k$ were found (as $k < 5$), where $K$ is between one and three orders of magnitude larger. This suggests a considerably more effective computation of the control law on the coarse-grained network, owing to the much smaller size of decision variables involved. Interestingly, the best reduction in the number of driver nodes, i.e., $\rho_D < 0.10$, occur in networks where the original density of driver nodes $K/N$ is considerable, i.e., over $0.40$. 

Metabolic and regulatory networks have been previously highlighted to be difficult to control because of the large ratios $K/N$, as occurs in the models of yeast and E. coli~\cite{Liu:2011vz}. Interestingly, here we find that for some of such models the coarse-grained networks turned out to be considerably more controllable. In particular the E. coli coarse-grained networks only require two driver macro-nodes instead of more than 300 nodes in the original networks. Ego networks reveal similar characteristics. Although they are extracted from different platforms, they are all difficult to control with $K/N > 0.95$. The corresponding coarse-grained networks, instead, show controllability ratios lower than $0.10$ in all cases. WWW networks exhibit $K/N$ values between $0.40$ and $0.76$ but the corresponding coarse-grained networks have similarly low values $k/n$ between $0.02$ and $0.04$.

\section{Conclusion}

In this work, we have introduced control equivalence, an optimality-preserving model reduction technique for linear control systems which can be computed in quasilinear time complexity in the number of nonzero matrix entries. We have conducted a large-scale evaluation on networks from public repositories, showing in particular that networks that are controlled by a minimal set of driver nodes often allow for further substantial optimality-preserving reductions. Future work will consider the optimality-preserving reduction of nonlinear control systems.

\bibliographystyle{IEEEtran}
\bibliography{cdc2023}

\appendix

\section*{Proofs}

\begin{proof}[Theorem~\ref{thm:main1}]
Let $L=L(\calH)$ be the aggregation matrix of the control equivalence $\calH$. As in Definition~\ref{def:red:sys}, we assume without loss of generality that $H_1,\ldots,H_k$ contain all at least one driver node, while $H_{k+1},\ldots,H_n$ have no driver nodes. 

For the proof of 1), pick any $u(\cdot) \in [m ; M]$ and note that
\begin{align*}
\partial_t L x(t) & = L \big(A x(t) + Bu(t)\big)
= L A \bar{L} L x(t) + LB u(t) \\
& = \hA L x(t) + LBu(t)
= \hA \hx(t) + \hB \hu(t) \\
& = \partial_t \hx(t),
\end{align*}
where the second identity holds because $\calH$ is a control equivalence, while the third and the last identity follow from the definition of the reduced model. Since the costs $F$ and $R$ are constant on $\calH$, the above calculation yields 1). 

Part 2) follows by picking any $\hu(\cdot) \in [\hm ; \hM]$, define $u(\cdot)$ using the formula in the part 2) and by reading the above calculation backwards. Specifically, the construction of $u(\cdot)$ from $\hu(t)$ ensures that $\hat{B}\hu(t) = L B u(t)$.
\end{proof}

\begin{proof}[Theorem~\ref{thm:main2}]
In the case $\calH$ is a control equivalence, Theorem~\ref{thm:main1} readily implies that the reduction is optimality-preserving. To show the converse, let us assume that the reduced model underlying a given partition $\calH$ is optimality-preserving for any final and running cost which are constant on $\calH$. Assume without loss of generality that $H_1,\ldots,H_k$ contain all at least one driver node, while $H_{k+1},\ldots,H_n$ have no driver nodes. 
We then fix arbitrary $T \geq 0$ and $u(\cdot) \in [m;M]$. Write $\| \cdot \|$ for the Euclidean norm and fix the final cost $F = 0$ that is obviously constant on $\calH$ and yields $\hat{F}=0$. Denoting by $x^u$ the solution of the original system induced by $u$, we consider the running cost
\begin{equation*}
R(t,x,v) = \| Lx - Lx^{u}(t) \| + \| LBv - LBu(t) \|
\end{equation*}
for any $v \in [m;M]$. It is immediate to see that $R$ is constant on $\calH$. As $R\big(t, x^{u}(t), u(t)\big) = 0$ for every $t \geq 0$, we have that
\begin{equation*}
0 = \int_0^{\T} R \big( t, x^{u}(t), u(t) \big) dt = J_{u}(x[0], \T ).
\end{equation*}
Moreover, $J_{ u' }(x[0], \T ) \geq 0$ for any $u'(\cdot) \in [m;M]$ as $R$ is non-negative, and so
\begin{equation*}
0 \leq V^{\inf} (x[0], \T) \leq J_{u}(x[0], \T ) = 0.
\end{equation*}
By assumption, $\hat{V}^{\inf} (\hx[0], \T)  = V^{\inf} (x[0], \T) = 0$. Since the optimal control is measurable by Filippov's theorem~\cite[Section 4.5]{Liberzon}, there thus exists $\hu(\cdot) \in [\hm;\hM]$ such that
\begin{equation*}
\hat{J}_{ \hu }(\hx[0],\T) = \hat{V}^{\inf}(\hx[0], \T) = 0 .
\end{equation*}
Writing explicitly the reduced value functional, 
the above equation gives
\begin{align*}
0 & = \hat{J}_{\hu}(\hx[0],\T) \\
& = \int_{0}^\T \big( \| \hx^{\hu}(t) - Lx^{u}(t) \| + \| \hB \hu(t) - LBu(t) \| \big) dt ,
\end{align*}
where $\hx^{\hu}(t)$ solves $\partial_t \hx^{\hu}(t) = \hA \hx^{\hu}(t) + \hB \hu(t)$. Hence, $\hx^{\hu}(t) - Lx^{u}(t)$ and $\hB \hu(t) - LBu(t)$ are zero almost everywhere on $[0;T]$. This and the fundamental theorem of calculus then imply that 
$\hx^{\hu} = Lx^{u}$ everywhere. 
Moreover, there exists a sequence of points $(t_\nu)_\nu$ converging to zero that satisfies for all $\nu \geq 0$:
\begin{align*}
\hA \hx^{\hu}(t_\nu) + \hB \hu(t_\nu) & = \partial_t \hx^{\hu}(t_\nu) \\
& = \partial_t Lx^u(t_\nu) \\
& = L A x^u(t_\nu) + LBu(t_\nu) ,
\end{align*}
yielding specifically $\hA \hx^{\hu}(t_\nu) = L A x^u(t_\nu)$ for all $\nu \geq 0$. Since this yields $\hA \hx[0] = LA x[0]$ and $x[0]$ was chosen arbitrarily, we infer that $\calH$ is a control equivalence.
\end{proof}

\begin{proof}[Theorem~\ref{thm:main3}]
Control equivalence can be seen as the extension of forward equivalence to control systems. Thanks to the fact that the algebraic description of control equivalence (i.e., $L A  = L A \bar{L} L$) and the formula for $\hA$ are as in the case of forward equivalence (and $\hB$ is known), the statement readily follows from~\cite{pnas17}.
\end{proof}

\end{document}